# Metasurfaces help lasers to mode-lock

Basudeb Sain and Thomas Zentgraf

Department of Physics, Paderborn University, 33098 Paderborn, Germany

**Metasurface saturable absorbers may result in versatile mode-locking that allows one to obtain stable ultrashort laser pulses with high repetition rates and peak powers, along with broadband operation, within fiber to solid-state laser cavities.**

Saturable absorption is a nonlinear optical effect where the material absorption coefficients decrease with increasing intensity of the incident light; therefore, the material becomes more transparent under intense light irradiation. Materials that show this kind of effect are called saturable absorbers (SAs). Saturable absorbers have become widely used as passive mode-locking elements for the generation of ultrashort laser pulses within a laser cavity. Although many materials possess saturable absorption properties (sometimes at very high optical intensities), a true saturable absorber is characterized by three main parameters: the modulation depth, the saturation intensity, and the nonsaturable losses. Semiconductor saturable absorber mirrors (SESAMs) are the most traditional SAs and are commonly used in solid-state and fiber lasers[1]. Ion-doped crystals are another type of SA frequently used in mode-locking or Q-switching applications[2]. However, these traditional SAs are wavelength-sensitive and suffer from a narrow modulation bandwidth. In recent years, two-dimensional (2D) materials such as graphene, transition-metal dichalcogenides (TMDs), black phosphorus, topological insulators, and two-dimensional transition metal carbides and nitrides (MXenes) have been extensively studied as alternatives to traditional SAs[3,4]. Despite their excellent saturable absorption properties, including an ultrafast recovery time, an acquiescent modulation depth, and wideband absorption, some shortcomings within the individual material category limit the fitness of 2D materials for their practical utilization in mode-locked lasers. For instance, graphene suffers from a low modulation depth and large nonsaturable losses; similarly, black phosphorus possesses high surface instability in the ambient atmosphere and may degrade quickly. Therefore, the quest is to find new ways to make saturable absorbers more versatile, dynamic, efficient and cost-effective. In this context, metasurfaces, which are blessed with arbitrary design freedom along with a large material selection, could become a superior alternative[5].

Metasurfaces consisting of subwavelength metallic/dielectric antennas (meta-atoms) can provide a versatile platform for the manipulation and control of light propagation and light-matter interactions[6]. Their uniqueness in the tailoring phase, amplitude, polarization, and spectral responses of light at the subwavelength scale with an exceptional ultrahigh spatial resolution has made them unprecedented candidates in modern nanophotonics research. By proper selection of the nanoantenna geometry (especially the anisotropic geometry under different polarization illumination), their spatial locations on a surface, and their material, it is possible to tailor their resonance behavior, implying an alteration of the transmission, reflection, absorption, and emission. In recent years, significant advancements in nanofabrication techniques have considerably broadened the experimental framework to address important issues such as wavefront shaping, upsurge in diffraction and conversion efficiency, and nonlinear phase matching in the field of nanophotonics[7-9]. As a result, metasurfaces have extended their functionalities in almost every part of linear and nonlinear optics, including quantum optics[10] and temporal pulse-shaping[11] applications.

During the past, plasmonic nanostructures were employed to study their nonlinear absorption sub-picosecond time-scale properties, such as hot electron dynamics, transient absorption, and saturable absorption, by manipulation of their surface plasmon resonances[12-14]. The results showed that a metasurface made of plasmonic nanostructures has the potential to be a saturable absorber for passive mode-locking in laser cavities.

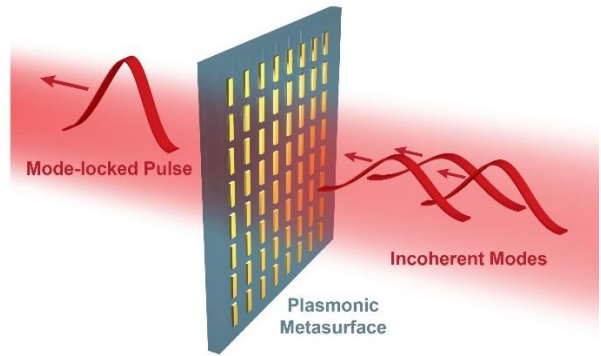

Figure 1. Schematic illustration of a plasmonic metasurface with periodically arranged gold nanorods for use as a saturable absorber. The fast saturation of the light absorption by the plasmonic nanorods results in coupling of the laser cavity modes, forming a short pulse.

In a recent publication, Wang et al. reported a successful demonstration of mode-locking by saturable absorbers using plasmonic metasurfaces, offering better modulation performances than 2D materials and traditional saturable absorbers[15]. For their realization, they integrated the metasurface saturable absorber into a fiber laser cavity to operate it in the pulsed regime. The metasurface SA was made of gold nanorods (NRs) with an optimized geometry arranged in a rectangular lattice. These small gold NRs provide for the polarization of light parallel to their long axes a localized surface plasmon polariton resonance that matches the wavelength of the laser input. To saturate the absorption of the NR array, they used a 1555 nm pulsed laser with a 500 ps pulse duration, a repetition rate of 100 kHz, and an average output power of 100 mW that corresponds to a peak power of 2 kW. In their experiments, they obtained a maximum modulation depth of ~60% at the corresponding plasmonic resonance for NRs with a saturation threshold of ~50 mW. The reported modulation depth is significantly higher than that presented in previous reports in the literature for all 2D materials, colloidal gold NRs and SESAMs. Based on the authors' results, it is implied that the nonlinear absorption has a direct link with the plasmonic resonances of the NR array. Hence, not only the resonances of the single NRs but also the coupling among them due to lattice resonances plays an important role. Furthermore, the authors integrated the saturable metasurface into a fiber laser cavity for ultrashort laser pulse generation. By using an overall cavity length of 7.5 m consisting of a standard single-mode silica fiber with an anomalous net chromatic dispersion, they achieved stable mode-locked regimes generating either ultrashort solitons or soliton molecules, demonstrating a 729 fs pulse train with a repetition rate of 28.2 MHz and a large signal-to-noise ratio of 75 dB.

Based upon the enormous potential of light manipulation by metasurfaces, the work by Wang et al. gives a glimpse of a possible application where plasmonic metasurfaces can improve the performance of a system. However, further research needs to address several important challenges related to metasurface SAs for laser mode-locking. These challenges correspond to the realization of metasurface SAs with a wide spectral range, high repetition rate, fast relaxation time of excited carriers, and high damage thresholds. All these properties are crucial for obtaining stable high-repetition ultrashort pulses with high peak power. The main limitations of plasmonic metasurfaces include high dissipative losses and inevitable thermal heating, leading in general to low optical damage thresholds. Some of the challenges might be solvable using dielectric or metal-dielectric hybrid metasurfaces. Metasurfaces made of dielectric nanostructures can support a larger variety of spatial modes (resonances) and have the ability to withstand much higher optical field intensities[16]. The excitation and interplay of multipolar resonance modes in these nanostructures might provide a route toward achieving the desired field enhancement for tailoring the spatial absorption to particular doped regions. In addition, a combination of metal-dielectric nanostructures might further improve functionalities in terms of strongly localized hybrid modes. We envision that metasurface SAs may become a versatile and cheap way to realize passive mode-locking and Q-switching in ultrashort pulsed fiber and solid-state laser systems.

**Acknowledgements**
This work has received funding from the European Research Council (ERC) under the European Union's Horizon 2020 research and innovation programme (grant agreement No 724306).